\pdfoutput=1
\documentclass[a4paper, amsfonts, amssymb, amsmath, reprint, showkeys, nofootinbib, twoside]{revtex4-1}
\usepackage[english]{babel}
\usepackage[utf8]{inputenc}
\usepackage[colorinlistoftodos, color=green!40, prependcaption]{todonotes}
\usepackage{amsthm}
\usepackage{mathtools}
\usepackage{physics}
\usepackage{xcolor}
\usepackage{graphicx}
\usepackage[left=23mm,right=13mm,top=35mm,columnsep=15pt]{geometry} 
\usepackage{adjustbox}
\usepackage{placeins}
\usepackage[T1]{fontenc}
\usepackage{lipsum}
\usepackage{csquotes}
\usepackage[pdftex, pdftitle={Article}, pdfauthor={Author}]{hyperref} % For hyperlinks in the PDF
\begin{document}
\title{Evaluating Deep Learning Models for Multiclass Classification of LIGO Gravitational Wave Glitches}

\author{Rudhresh Manoharan}
    \email[Corresponding author: ]{Rudhresh_Manoharan1@baylor.edu}
    \affiliation{EUCOS-CASPER, Physics Department, Baylor University, Waco, TX 76798-7316, USA}

\author{Gerald B. Cleaver}
    \email{Gerald_Cleaver@baylor.edu}
    \affiliation{EUCOS-CASPER, Physics Department, Baylor University, Waco, TX 76798-7316, USA}
    
\date{\today} % Leave empty to omit a date

\begin{abstract}
\noindent
Gravitational-wave detectors are affected by short-duration non-Gaussian noise transients, commonly referred to as glitches, which can obscure astrophysical signals and complicate downstream analyses. While recent work has demonstrated the effectiveness of deep learning models for glitch classification using image-based time–frequency representations, comparatively less attention has been paid to systematic evaluations of machine-learning architectures operating directly on tabular glitch metadata. In this work, we present a comprehensive benchmark of classical and deep learning models for multiclass glitch classification using numerical features derived from the Gravity Spy dataset. We compare gradient-boosted decision trees with a diverse set of neural architectures, including multilayer perceptrons, attention-based models, and neural decision ensembles, and evaluate them in terms of classification performance, inference efficiency, parameter efficiency, data-scaling behavior, and cross-model interpretability alignment. We find that while tree-based methods remain strong baselines for tabular data, several deep learning models achieve competitive performance with substantially fewer parameters and exhibit distinct inductive biases and scaling behavior. Furthermore, a cross-model attribution analysis reveals that different neural architectures recover partially consistent feature-importance hierarchies, providing new insight into interpretability structure across models. These results clarify trade-offs between performance, complexity, data efficiency, and interpretability in tabular gravitational-wave analyses and provide practical guidance for deploying machine-learning models in detector characterization pipelines.
\end{abstract}


\maketitle

\section{Introduction} \label{sec:introduction}
\noindent
The direct detection of gravitational waves marked the beginning of gravitational-wave astronomy and enabled precision studies of compact-object populations and fundamental physics \cite{abbott2016observation}. Since then, successive observing runs by the LIGO–Virgo–KAGRA (LVK) collaboration have produced increasingly large catalogs of gravitational-wave events \cite{abac2025gwtc}. Achieving and maintaining the sensitivity required for these discoveries, however, necessitates careful characterization of detector noise, particularly transient non-Gaussian artifacts known as glitches.

Machine learning has become an integral component of modern gravitational-wave data analysis, with applications spanning detection, parameter estimation, detector characterization, and noise mitigation \cite{cuoco2020enhancing, cuoco2025applications}. A prominent example is the Gravity Spy project, which combines citizen science with supervised machine learning to classify glitches based on their morphology \cite{zevin2017gravity}. Over multiple observing runs, Gravity Spy has evolved into both a widely used dataset and a framework for studying detector noise, culminating in curated datasets for O3 \cite{glanzer2023data} and advanced classifiers designed for O4 \cite{wu2025advancing}, where O3 and O4 denote the third and fourth observing runs of the Advanced LIGO–Virgo–KAGRA collaboration. These efforts have enabled the discovery of new glitch classes and informed detector commissioning strategies \cite{zevin2024gravity, soni2021discovering}.

Much of the prior work on glitch classification has focused on image-based representations, leveraging convolutional neural networks, transfer learning, and vision transformers applied to spectrograms \cite{PhysRevD.97.101501, fernandes2023convolutional, razzano2018image, bahaadini2017deep}. Generative models have also been explored for data augmentation and simulation \cite{yan2022improving, powell2023generating}. While these approaches achieve high accuracy, they rely on time–frequency representations and substantial preprocessing pipelines.

Beyond supervised image-based pipelines, recent studies have explored a broader range of machine-learning strategies for glitch characterization, including unsupervised learning, uncertainty-aware classification, and alternative feature representations. Unsupervised and self-supervised approaches have been applied to cluster and characterize transient noise without predefined labels, enabling discovery-driven analyses in both LIGO and KAGRA data \cite{oshino2025glitch, li2024cross}. Vision-transformer architectures and attention-based models have been investigated to capture long-range structure in transient noise, while conformal prediction has been introduced to quantify classification uncertainty in safety-critical settings \cite{srivastava2025vision, malz2025classification}. Complementary work has emphasized interpretability and physically motivated diagnostics for transient classification---for example, ``waveform surgery''-style attribution analyses and related interpretation tools---alongside representation choices such as wavelet scattering transforms \cite{PhysRevD.109.022006, licciardi2025wavelet}. Together, these efforts highlight the growing methodological diversity of machine-learning tools for gravitational-wave detector characterization.

In contrast, many gravitational-wave analyses produce rich tabular metadata describing glitch properties, auxiliary channel correlations, and signal-consistency statistics. Despite the widespread availability of such structured data, comparatively fewer studies have systematically evaluated deep learning models designed specifically for tabular inputs in the context of gravitational-wave glitches. Recent work in machine learning has shown that tabular data pose unique challenges, with tree-based ensembles often outperforming deep neural networks unless architectural inductive biases are carefully matched to the data \cite{ye2024closer}.

Motivated by this gap, we present a focused benchmark of classical and deep learning models for multiclass glitch classification using tabular numerical features from the Gravity Spy dataset. Our goal is not to propose a new architecture, but rather to clarify the empirical trade-offs between performance, efficiency, complexity, and scalability across a representative set of models. By grounding our analysis in realistic detector metadata and emphasizing interpretable evaluation metrics, we aim to provide actionable guidance for future machine-learning deployments in gravitational-wave detector characterization. The full reproducible implementation of this work is publicly available, with a versioned archival release provided via Zenodo.  %I believe leaving the sections in separate files is more organized, change it if you desire 
\section{Dataset and Features} \label{sec:dataset}
\noindent
We use the Gravity Spy O3 dataset \cite{glanzer2023data}, which provides labeled examples of transient noise artifacts identified in Advanced LIGO data. Each example corresponds to a detector glitch and is annotated with a discrete class label as well as a set of numerical features derived from time–frequency and signal-consistency analyses. The dataset contains multiple glitch classes with pronounced class imbalance, reflecting realistic operating conditions of ground-based gravitational-wave detectors.

To investigate data-scaling behavior, we construct two stratified datasets: a sampled subset containing $5\times10^{4}$ examples and a larger dataset containing approximately $5\times10^{5}$ examples. In both cases, stratification preserves the original class proportions. All experiments are framed as a multiclass classification problem, and performance is evaluated using the weighted F1 score to account for class imbalance.

\begin{figure}[t]
  \centering
  \includegraphics[width=\linewidth]{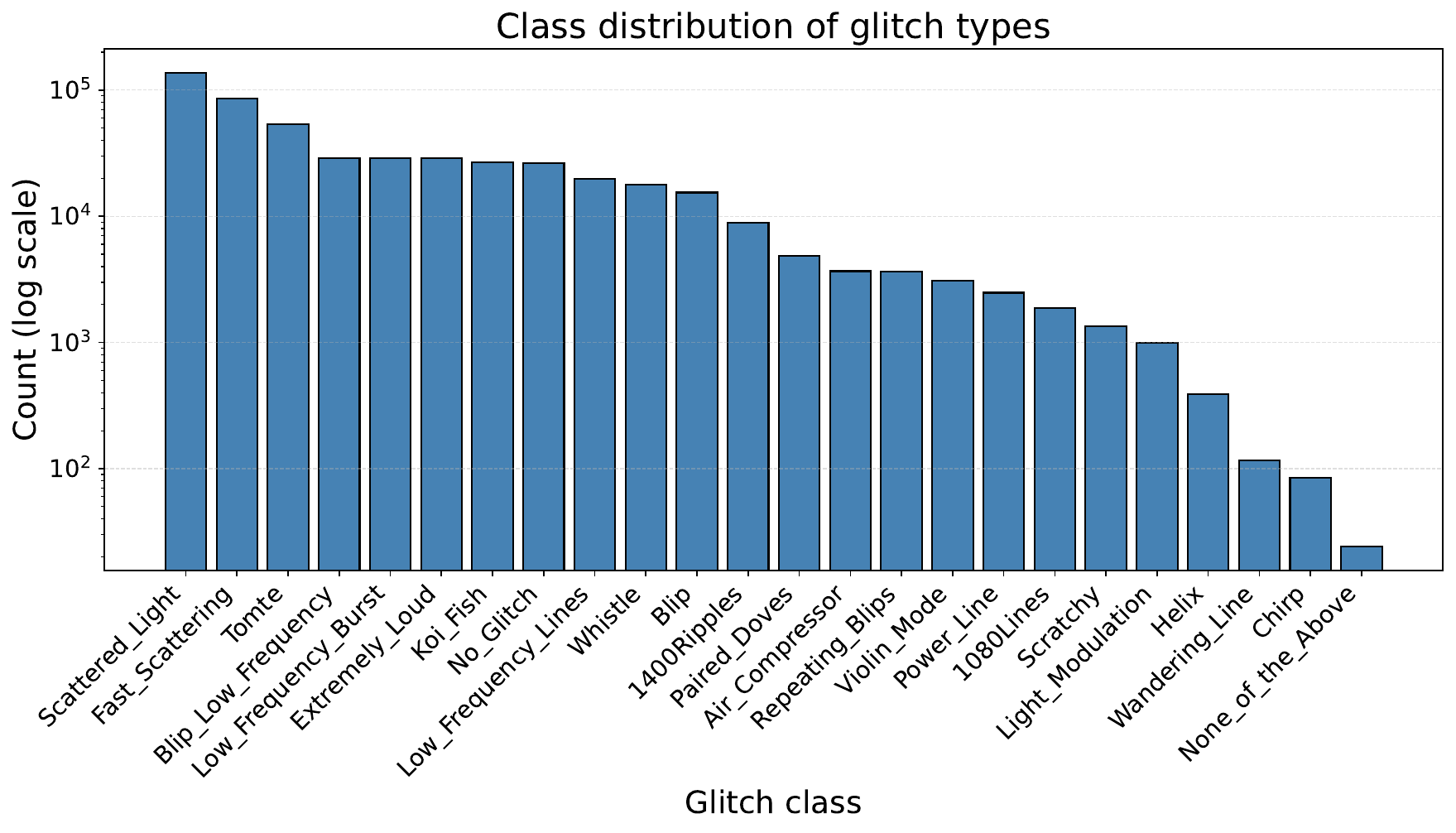}
  \caption{
  \textbf{Class distribution of glitch types in the Gravity Spy O3 dataset.}
  The histogram shows the number of examples per glitch class on a logarithmic scale for the full dataset ($\sim500\,000$ examples). The wide dynamic range highlights the severe class imbalance characteristic of real gravitational-wave detector data.
  }
  \label{fig:dataset_overview}
\end{figure}

Figure~\ref{fig:dataset_overview} shows the class distribution of glitch types in the full Gravity Spy O3 dataset on a logarithmic scale, highlighting the pronounced class imbalance across glitch categories. This imbalance reflects realistic detector conditions and motivates the use of weighted evaluation metrics throughout this study.

The models are trained using a compact set of nine numerical features that characterize the temporal, spectral, and amplitude properties of each glitch. These features, summarized in Table \ref{tab:features}, are derived directly from detector metadata and signal-processing pipelines and retain clear physical interpretation.

\begin{table}[t]
\centering
\caption{Numerical features used for tabular glitch classification and their physical interpretation.}
\label{tab:features}
\begin{tabular}{p{0.27\linewidth} p{0.65\linewidth}}
\hline
\textbf{Feature} & \textbf{Physical meaning / unit} \\
\hline
peak\_time & GPS second of maximum glitch amplitude (s) \\
peak\_time\_ns & Nanosecond offset within the GPS second (ns) \\
start\_time\_ns & Nanosecond timestamp at glitch onset (ns) \\
duration & Temporal extent of the glitch waveform (s) \\
peak\_frequency & Frequency of maximum signal power (Hz) \\
central\_freq & Mean spectral centroid of the glitch (Hz) \\
amplitude & Measured strain amplitude (dimensionless) \\
snr & Signal-to-noise ratio of the glitch (dimensionless) \\
q\_value & Quality factor characterizing spectral coherence \\
\hline
\end{tabular}
\end{table}

\section{Models and Training Procedure} \label{sec:methods}
\noindent
We evaluate a diverse set of machine-learning models commonly used for supervised learning on tabular data. As a strong classical baseline, we include gradient-boosted decision trees implemented via XGBoost, which are known to perform robustly on structured numerical features and frequently outperform deep neural networks in tabular settings. 

The neural models considered in this study include a multilayer perceptron (MLP), TabNet \cite{arik2021tabnet}, TabTransformer \cite{huang2020tabtransformer}, FT-Transformer \cite{gorishniy2021revisiting}, AutoInt \cite{song2019autoint}, DANet \cite{mia2023danet}, NODE \cite{popov2019neural}, GATE \cite{joseph2022gandalf}, and GANDALF \cite{joseph2022gandalf}. These architectures span a broad range of inductive biases, including feature-wise attention mechanisms, sequential decision steps, neural decision ensembles, and transformer-based feature interactions. Collectively, this model suite enables a systematic comparison between classical ensemble methods and modern deep learning approaches tailored to tabular data.

All neural models are implemented and trained using the \texttt{PyTorch Tabular} framework \cite{joseph2021pytorchtabularframeworkdeep}, which provides a unified API for state-of-the-art tabular deep learning architectures built on PyTorch and PyTorch Lightning. This framework allows consistent data handling, training procedures, and evaluation protocols across models, reducing implementation-induced variability in the benchmark.

For each dataset configuration, the data are split into training, validation, and test sets using stratified sampling to preserve class proportions, with a fixed split of $64\%/16\%/20\%$ for training, validation, and testing, respectively. 
Hyperparameters are optimized using the Optuna framework, with $100$ trials per model.
For each trial, performance is evaluated using stratified $3$-fold cross-validation on the training set, and the model configuration achieving the highest cross-validated weighted F1 score is selected.
Early stopping is employed based on validation performance to prevent overfitting.

To ensure fair comparison, training procedures and evaluation metrics are held fixed across datasets and architectures.
Following hyperparameter selection, each model is retrained using $15$ independent random seeds, and all reported results correspond to the mean weighted F1 score evaluated on the held-out test set.
All reported test-set results correspond to the best-performing model checkpoint selected according to validation weighted F1 score.

\begin{figure}[t] 
\centering 
\includegraphics[width=\linewidth]{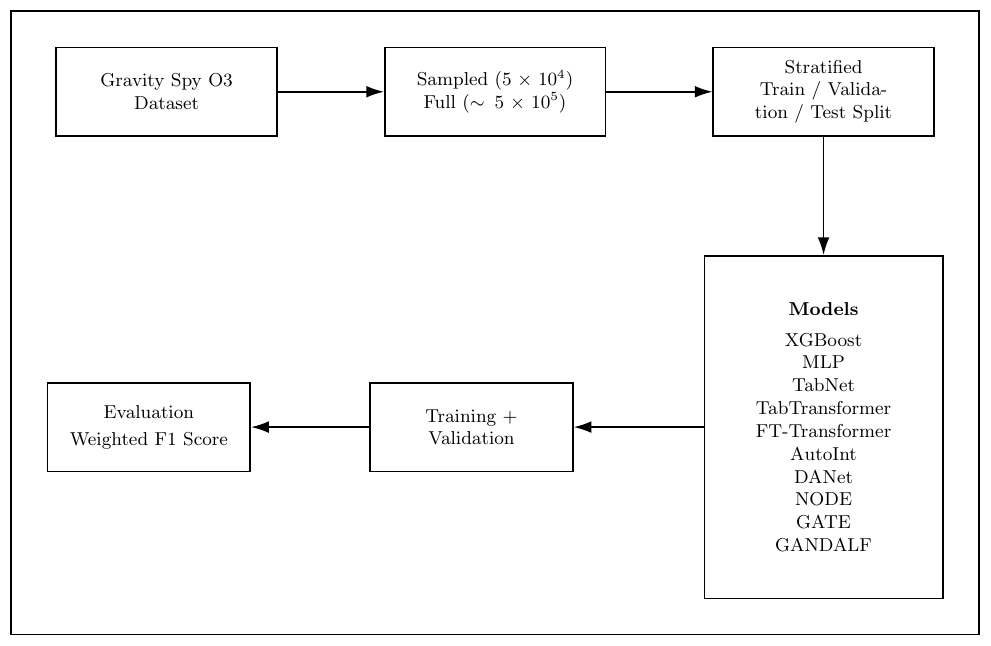} \caption{ 
\textbf{Overview of the modeling and evaluation pipeline.} The Gravity Spy O3 dataset is partitioned into sampled ($5\times10^4$) and full ($\sim5\times10^5$) configurations, each followed by a stratified train/validation/test split.
A suite of classical and deep tabular models is trained under a unified protocol, with performance evaluated using the weighted F1 score to account for class imbalance. 
} 
\label{fig:workflow_overview} 
\end{figure}

Figure~\ref{fig:workflow_overview} provides a schematic overview of the end-to-end workflow used in this study, from dataset construction through model training and evaluation. The figure summarizes the data-splitting strategy, the set of models evaluated, and the unified training and validation protocol, serving as a visual reference for the comparative analyses presented in the following sections.

\section{Results} \label{sec:results}
\noindent
\subsection{Classification Performance and Statistical Robustness}
\label{subsec:classification_performance}

\begin{figure}[h]
\centering
\includegraphics[width=\linewidth]{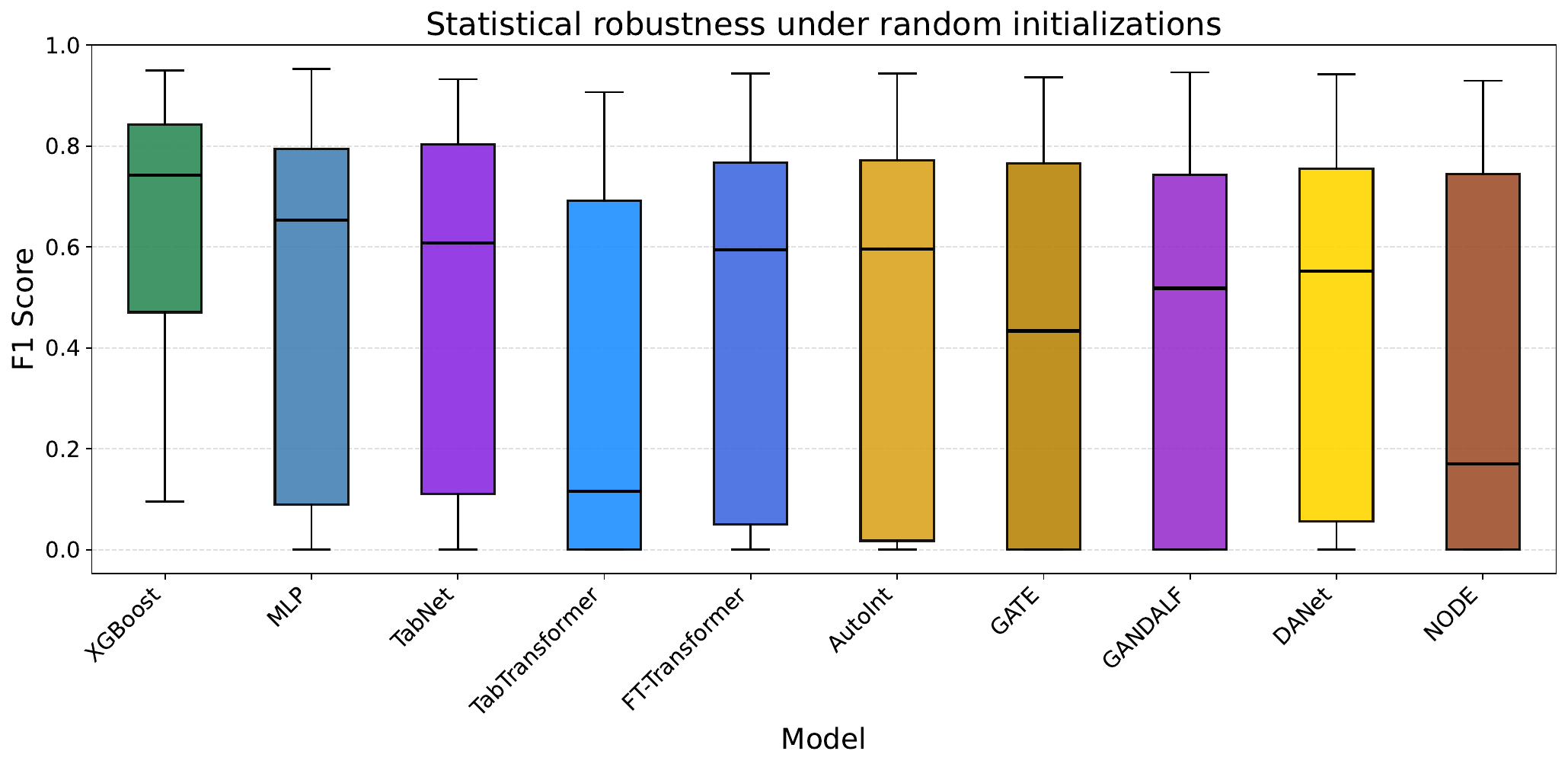}
\caption{\textbf{Distribution of weighted F1 scores across models.}}
\label{fig:boxplot_f1}
\end{figure}

\noindent
Figure~\ref{fig:boxplot_f1} summarizes the statistical robustness of all evaluated models through the distribution of weighted F1 scores obtained across multiple random initializations and cross-validation folds. Rather than reporting a single best-run score, this analysis characterizes the stability of each model under stochastic training effects, which is particularly important for deep learning architectures.

XGBoost achieves the highest median F1 score with a relatively narrow interquartile range, indicating both strong performance and high stability. This behavior is consistent with the well-established effectiveness of tree-based ensembles on tabular data. Among neural models, several architectures—including MLP, AutoInt, and GANDALF—exhibit competitive median performance, though with noticeably broader distributions, reflecting increased sensitivity to initialization and optimization dynamics.

Other models show substantially larger variance across runs, with some exhibiting long tails toward low F1 values. This highlights the importance of reporting performance distributions rather than single-point estimates, as mean or best-case metrics alone may obscure instability in practical deployments. Overall, Figure~\ref{fig:boxplot_f1} demonstrates that while deep tabular models can approach tree-based performance, their reliability varies significantly across architectures, motivating the subsequent analysis of efficiency, interpretability, and scaling behavior.

\subsection{Performance vs.\ Training Time}
\label{subsec:training_time}

\begin{figure}[t]
\centering
\includegraphics[width=\linewidth]{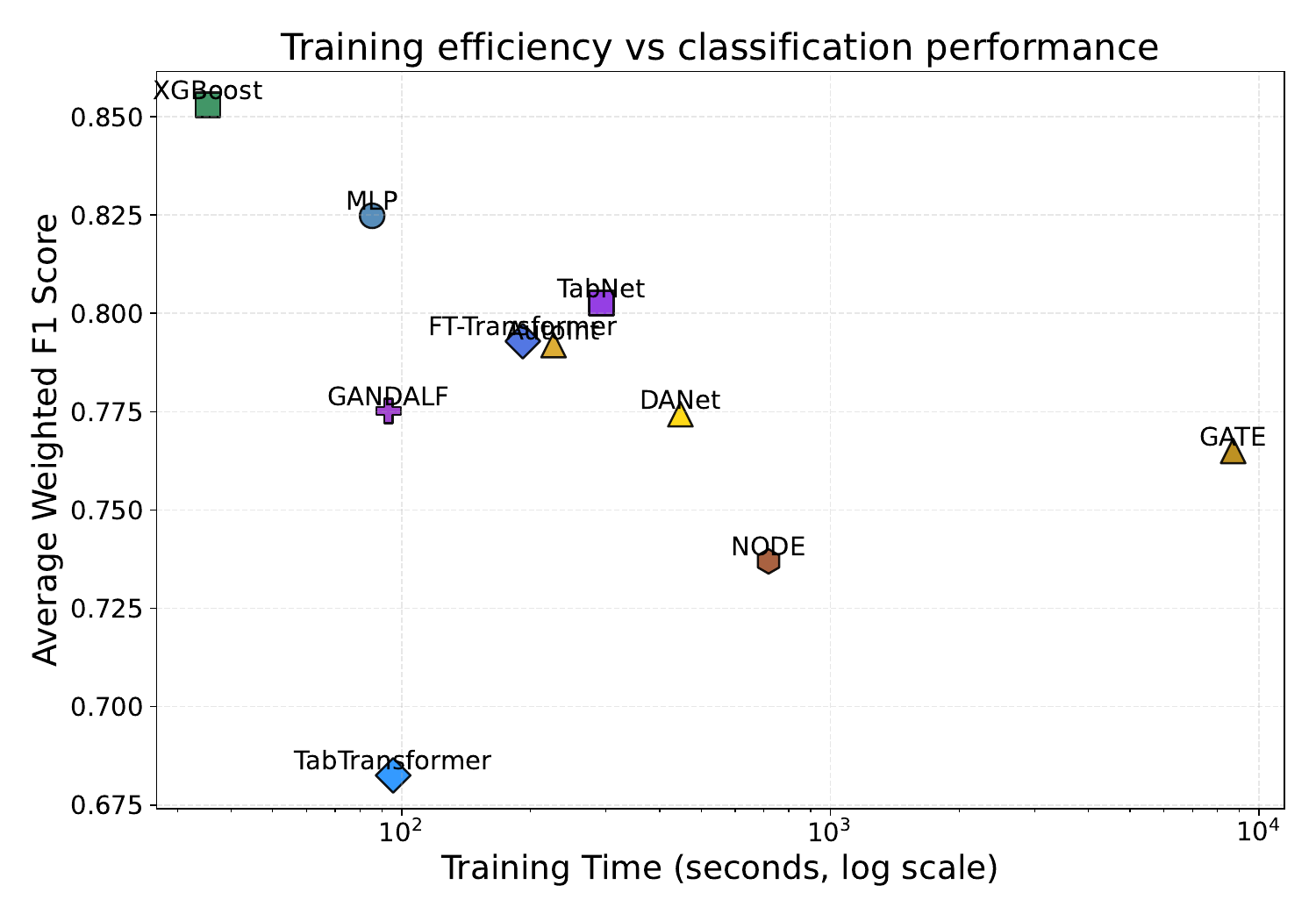}
\caption{\textbf{Weighted F1 score versus wall-clock training time for all models.} Training time is shown on a logarithmic scale to highlight differences spanning multiple orders of magnitude.}
\label{fig:f1_vs_training}
\end{figure}

\noindent
Figure~\ref{fig:f1_vs_training} compares the average weighted F1 score achieved by each model against the corresponding wall-clock training time, shown on a logarithmic scale. This comparison addresses the practical question of how computationally expensive it is to reach a given level of classification performance under a fixed training protocol.

Tree-based boosting achieves high performance with comparatively modest training cost, reflecting its efficiency on structured data. Neural models span a wide range of training times, with some architectures requiring substantially greater computational effort to attain competitive performance. Notably, several neural models achieve similar F1 scores despite order-of-magnitude differences in training time, underscoring the importance of architectural and optimization choices beyond raw capacity.

We emphasize that the training times shown correspond to wall-clock time required to fit a single model using the selected hyperparameters, including validation and early stopping, but excluding the total hyperparameter search budget. While trained models may be reused indefinitely at inference time, training cost remains relevant for model development, retraining on updated datasets, and resource allocation in large-scale or shared computing environments.

\subsection{Performance vs.\ Inference Time}
\label{subsec:inference_time}

\begin{figure}[t]
\centering
\includegraphics[width=\linewidth]{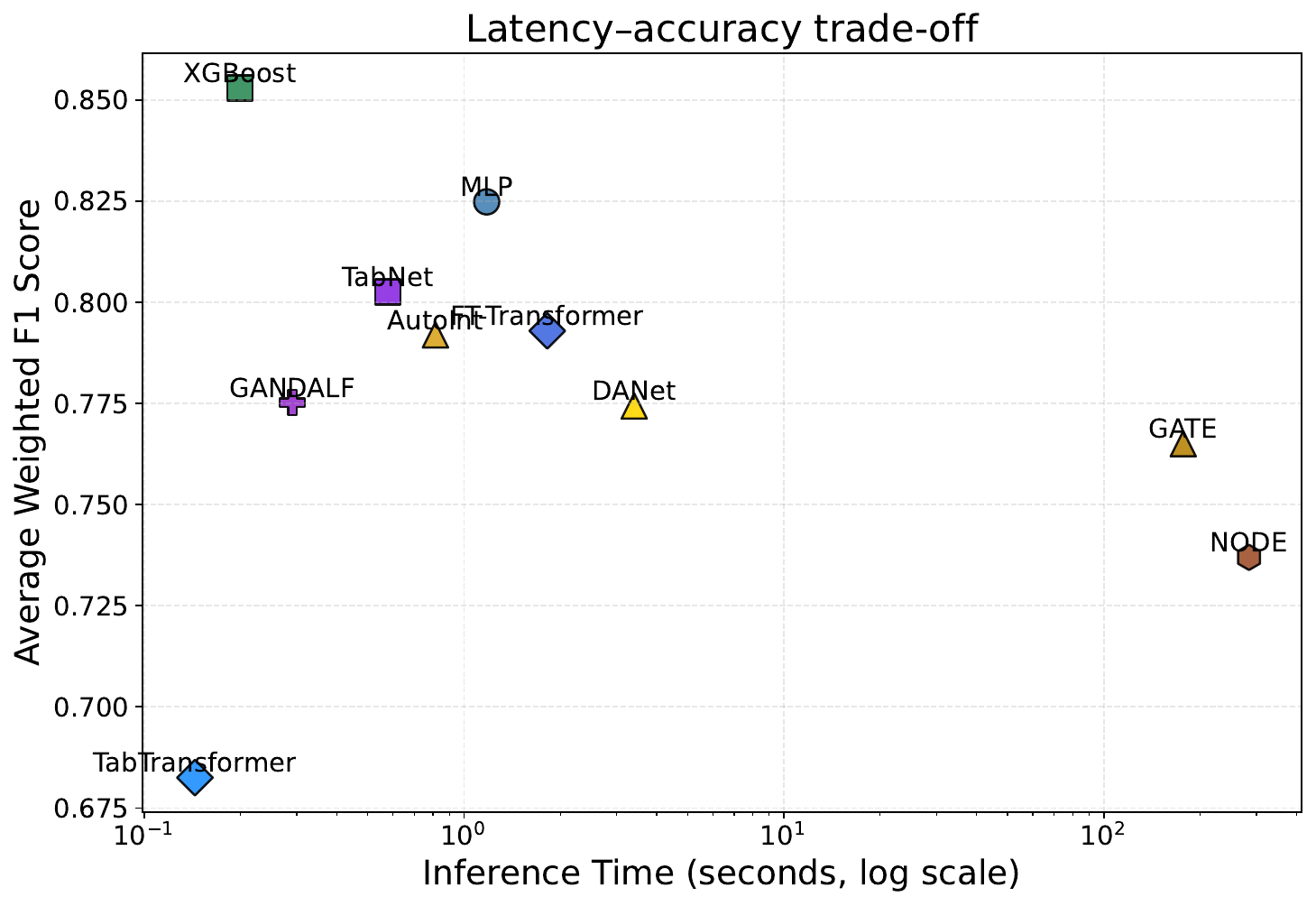}
\caption{\textbf{Weighted F1 score versus per-sample inference time for all models.} Inference time is shown on a logarithmic scale to emphasize differences relevant for low-latency deployment.}
\label{fig:f1_vs_inference}
\end{figure}

\noindent
Figure~\ref{fig:f1_vs_inference} compares the average weighted F1 score achieved by each model against the corresponding per-sample inference latency, shown on a logarithmic scale. Inference latency was measured as the end-to-end wall-clock time required to generate predictions using \texttt{TabularModel.predict} on a fixed stratified subset of the test set ($N=10112$), using identical batch size and hardware settings across all models. The reported per-sample latency is obtained by dividing the total prediction time by $N$; this measurement includes internal preprocessing, batching, and output formatting overhead in addition to the model forward pass. Unlike training cost, inference time cannot be amortized and directly impacts the feasibility of deployment in low-latency gravitational-wave data analysis pipelines.

The gradient-boosted decision tree baseline achieves strong performance with relatively low inference cost, reflecting its efficiency for structured feature representations. Several neural models exhibit competitive accuracy while maintaining inference times within an order of magnitude of the tree-based baseline, indicating that deep tabular architectures can be viable for near-real-time applications when appropriately selected.

Differences in inference latency arise from architectural design choices. Tree ensembles require traversal across many decision paths, with cost scaling with the number of trees and effective leaf structure. Attention-based tabular models introduce feature-interaction layers whose computational cost grows with embedding dimension, number of heads, and attention blocks. Sequential feature-selection architectures (e.g., masked decision-step models) may incur multiple internal passes per sample, while neural ensemble-style methods aggregate outputs across multiple learned components. These structural differences help explain why models achieving similar classification performance can nevertheless exhibit order-of-magnitude variation in inference latency.

At the same time, substantial variation is observed across neural architectures, with some models incurring significantly higher inference latency without commensurate gains in classification performance. These results highlight inference efficiency as an essential axis of comparison, complementary to training cost and peak accuracy, particularly for operational settings where rapid response and continuous processing are required.

\subsection{Performance vs.\ Model Complexity}
\label{subsec:model_complexity}

\begin{figure}[t]
\centering
\includegraphics[width=\linewidth]{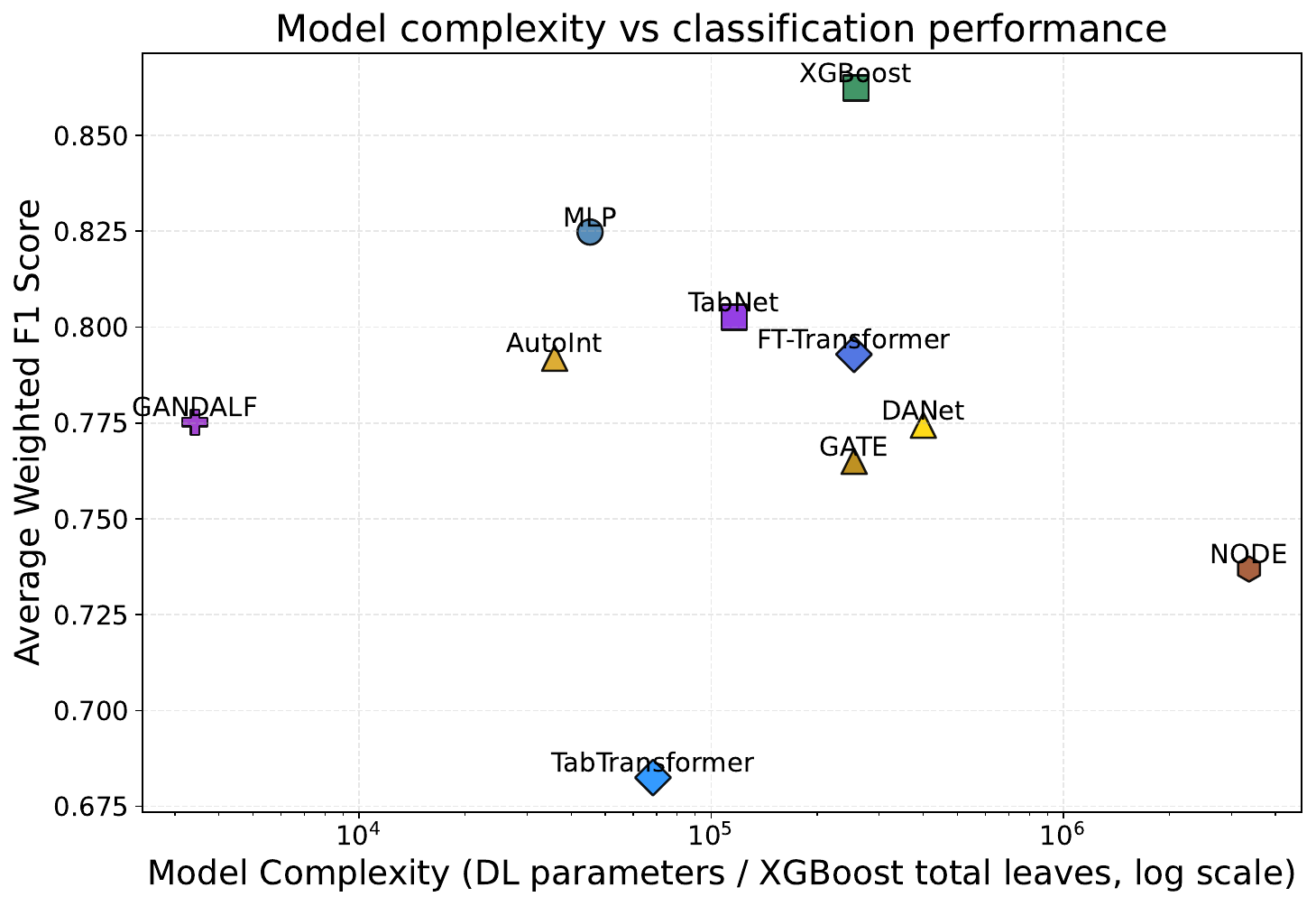}
\caption{\textbf{Weighted F1 score versus model complexity proxy for the full Gravity Spy O3 dataset.} Model complexity is measured by total trainable parameters for neural models and total leaf count for XGBoost, shown on a logarithmic scale.}
\label{fig:f1_vs_params}
\end{figure}

\noindent
To assess empirical parameter efficiency, Figure~\ref{fig:f1_vs_params} shows the weighted F1 score as a function of model complexity. For neural models, complexity is measured by the total number of trainable parameters, while for XGBoost we report the total number of leaf nodes across all trees as an effective proxy for model capacity. All results correspond to evaluation on the full ($\sim500\,000$ examples) Gravity Spy O3 dataset.

The tree-based baseline achieves the highest overall performance at relatively high effective capacity, reflecting its strong inductive bias for structured tabular data. Among neural architectures, substantial variation is observed in how predictive performance scales with model size. Several models achieve competitive accuracy with orders of magnitude fewer parameters, indicating high parameter efficiency driven by architectural inductive bias rather than raw capacity.

In contrast, increasing model complexity does not uniformly translate into improved performance. Larger architectures exhibit clear saturation behavior, and in some cases reduced efficiency, highlighting the limits of over-parameterization for this dataset. Overall, these results demonstrate that performance, efficiency, and complexity are distinct axes of comparison for tabular glitch classification, and that careful architectural choice can yield strong performance without excessive model size.

\subsection{Feature Alignment Across Models}

\noindent
To assess whether different models rely on similar physical information when making predictions, we quantify the alignment of feature-importance rankings across models. For each neural architecture, we compute the Spearman rank correlation between its feature-importance vector and that obtained from XGBoost, which serves as a well-established and interpretable baseline for tabular data. Feature importance for neural models is derived using Captum, while TreeSHAP is used for XGBoost.

\begin{figure}[h]
\centering
\includegraphics[width=\linewidth]{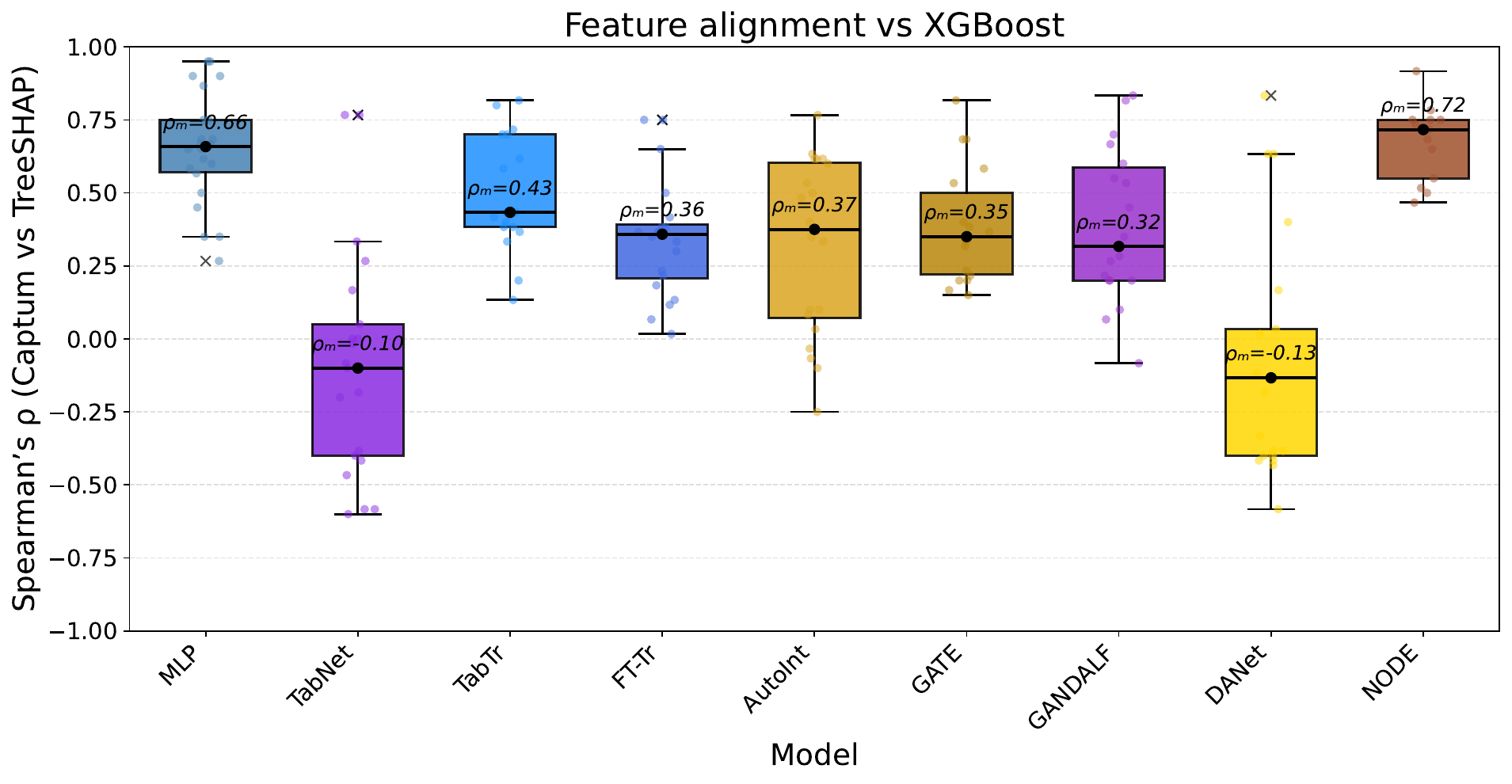}
\caption{\textbf{Spearman rank correlation between feature-importance vectors of each model and XGBoost, computed across glitch classes.} Boxes indicate the distribution of correlations, with median values annotated.}
\label{fig:feature_alignment}
\end{figure}

Figure~\ref{fig:feature_alignment} shows the distribution of Spearman correlation coefficients across glitch classes for each model. The results reveal substantial heterogeneity in interpretability alignment. Among the neural models, NODE exhibits the strongest alignment with XGBoost, achieving the highest median correlation ($\rho_m = 0.72$), indicating that its learned feature rankings closely track those of the tree-based model. MLP also shows relatively strong agreement ($\rho_m = 0.66$), despite its architectural simplicity.

Several transformer-based and attention-driven models exhibit moderate positive alignment, including TabTransformer ($\rho_m = 0.43$), FT-Transformer ($\rho_m = 0.36$), AutoInt ($\rho_m = 0.37$), GATE ($\rho_m = 0.35$), and GANDALF ($\rho_m = 0.32$). These models partially recover the feature ordering learned by XGBoost, suggesting that they capture some of the same physical structure while also introducing model-specific inductive biases.

In contrast, TabNet and DANet display weak or negative median correlations ($\rho_m = -0.10$ and $\rho_m = -0.13$, respectively), indicating that their internal feature attributions diverge substantially from those of the tree-based baseline. Unlike static tree ensembles, TabNet employs sequential decision steps with learned sparse attention masks that dynamically select subsets of features at each stage of inference. This step-wise masking mechanism can emphasize conditional feature usage and higher-order interactions that may not be reflected in global feature-ranking summaries. Similarly, DANet incorporates adaptive feature selection modules and dense connectivity patterns that promote interaction-driven representations across layers. These architectural mechanisms can redistribute attribution mass across interacting features, leading to global ranking structures that differ from those recovered by TreeSHAP in gradient-boosted trees. The observed divergence therefore likely reflects differences in how these models encode feature interactions rather than purely stochastic variability.

Overall, these results demonstrate that high classification performance does not necessarily imply interpretability alignment across models. Feature-ranking consistency emerges as an independent axis of comparison, highlighting that different architectures may achieve similar predictive accuracy while relying on distinct subsets or orderings of physical features.

\subsection{Cross-Model Interpretability Heatmap}

\begin{figure}[h]
\centering
\includegraphics[width=\linewidth]{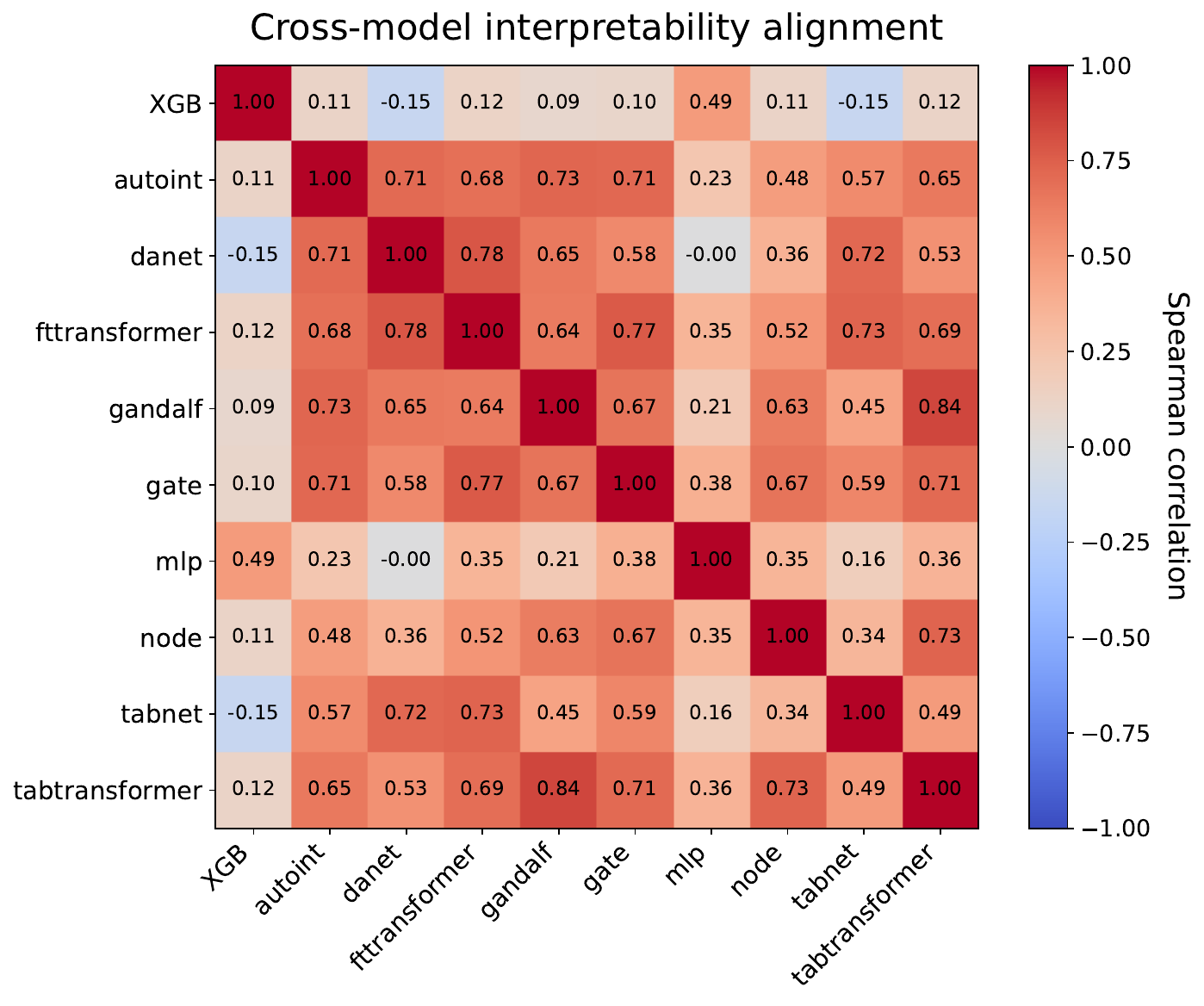}
\caption{\textbf{Cross-model interpretability alignment measured via Spearman rank correlations between flattened per-class feature-importance vectors.} Each entry quantifies the agreement between two models in how they rank tabular features across all glitch classes. Clusters of high correlation indicate models with similar inductive biases and attribution structure, while lower correlations highlight divergent feature-selection behavior.}
\label{fig:interpretability_heatmap}
\end{figure}

\noindent
To further examine how different models encode feature-importance structure, we compute pairwise Spearman rank correlations between flattened per-class feature-importance vectors across all models. Figure~\ref{fig:interpretability_heatmap} presents the resulting cross-model correlation matrix, providing a global view of interpretability alignment beyond comparisons to a single baseline.
To our knowledge, this constitutes the first systematic, quantitative study of cross-model interpretability alignment for tabular gravitational-wave glitch classification, extending beyond single-model explanations to a comparative, architecture-level analysis.

Several notable patterns emerge. Models sharing similar inductive biases exhibit strong mutual alignment: attention-based architectures such as TabTransformer, FT-Transformer, AutoInt, GATE, and GANDALF form a coherent cluster with consistently high correlations ($\rho \gtrsim 0.6$–$0.8$). This indicates that, despite architectural differences, these models emphasize similar subsets of features across glitch classes. NODE also aligns moderately well with this group, suggesting that neural decision-ensemble methods can converge toward comparable feature-importance structures.

In contrast, XGBoost displays comparatively weaker alignment with most deep learning models, reflecting fundamental differences between tree-based split criteria and gradient-based attribution mechanisms. The multilayer perceptron (MLP) occupies an intermediate position, showing partial alignment with both XGBoost and attention-based models, consistent with its lack of explicit feature-selection inductive bias.

Overall, this heatmap reveals that interpretability agreement is not uniform across models, but instead clusters according to architectural principles. These results complement the one-to-one alignment analysis in Figure~\ref{fig:feature_alignment}, demonstrating that multiple deep tabular models independently recover similar feature hierarchies, even when their absolute performance differs.
Such cross-model convergence has not previously been examined in the context of gravitational-wave detector characterization and provides new evidence that learned representations capture physically meaningful structure rather than model-specific artifacts.

\subsection{Class-Level Performance of the Best Model}

\begin{figure}[h]
\centering
\includegraphics[width=\linewidth]{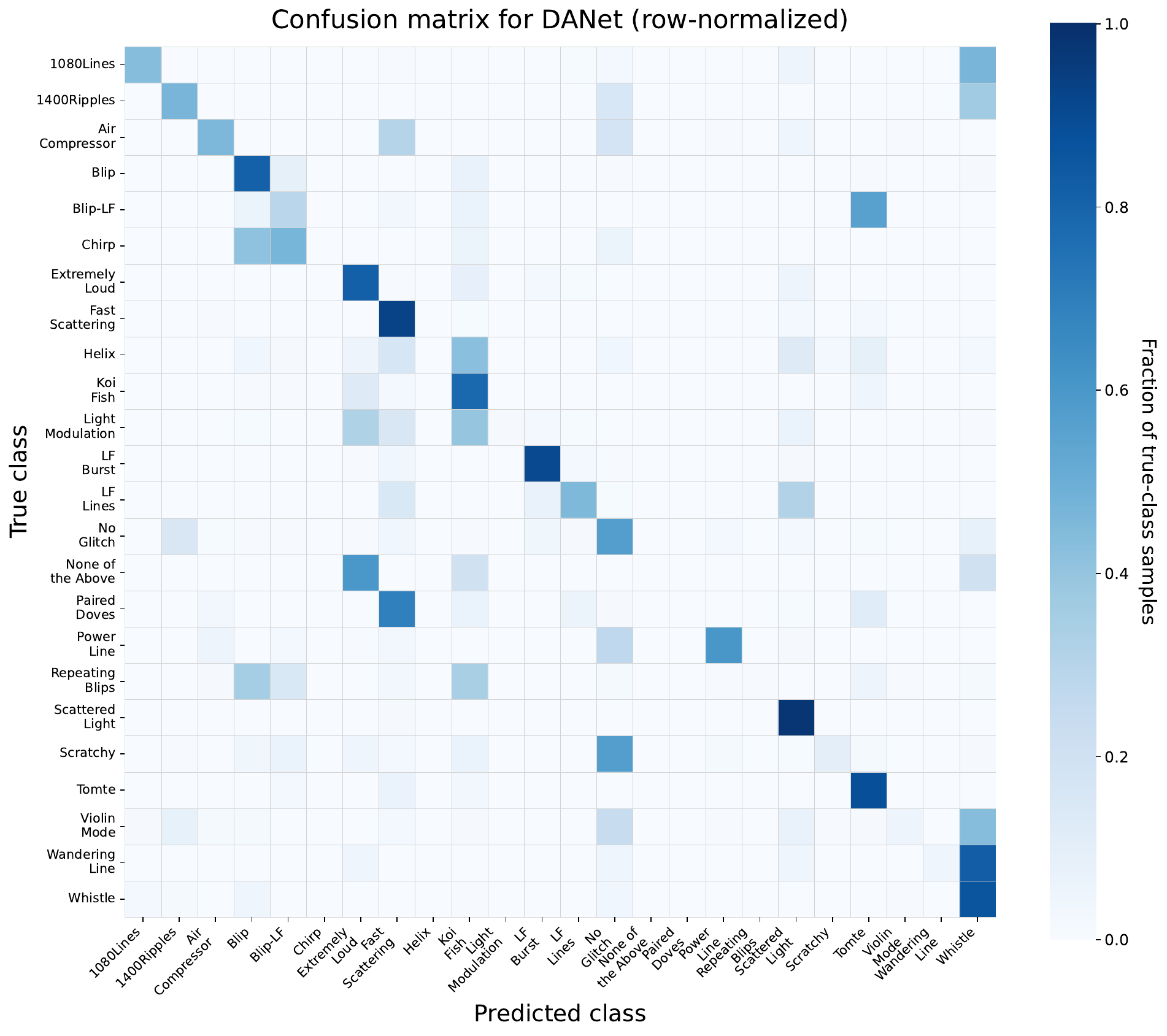}
\caption{\textbf{Row-normalized confusion matrix for DANet on the 24-class glitch classification task.} Each row is normalized to unity, so entries represent the fraction of samples from a given true class assigned to each predicted class. Strong diagonal structure indicates effective class separation, while off-diagonal entries reveal systematic confusions between morphologically similar glitch types. The normalization highlights class-specific performance independent of dataset imbalance, enabling direct comparison of recall across rare and dominant classes.}
\label{fig:confusion_normalized}
\end{figure}

\noindent
To further understand the behavior of the best-performing model, we examine the class-level prediction structure using a confusion matrix for DANet evaluated on the test set. The row-normalized confusion matrix (Figure~\ref{fig:confusion_normalized}) highlights systematic misclassifications between specific glitch classes while removing dependence on class frequency.

Overall, the matrix exhibits a strong diagonal structure, indicating robust class separation for several dominant glitch types such as Scattered Light, Fast Scattering, Low-Frequency Burst, and Tomte. However, notable off-diagonal patterns reveal consistent confusions among morphologically similar classes. In particular, Blip Low Frequency is frequently misclassified as Tomte, suggesting overlap in their time–frequency characteristics. Similarly, Air Compressor events are often confused with Fast Scattering and No Glitch, indicating that their spectral signatures may not be sufficiently distinctive in the current feature space. The model also struggles with rare or weakly represented classes such as Chirp, None of the Above, and Light Modulation, which exhibit low recall and are dispersed across multiple predicted categories. Additional confusion is observed between Repeating Blips and Blip, as well as between Wandering Line and Whistle, reflecting similarities in their underlying signal morphology.

\begin{figure}[h]
\centering
\includegraphics[width=\linewidth]{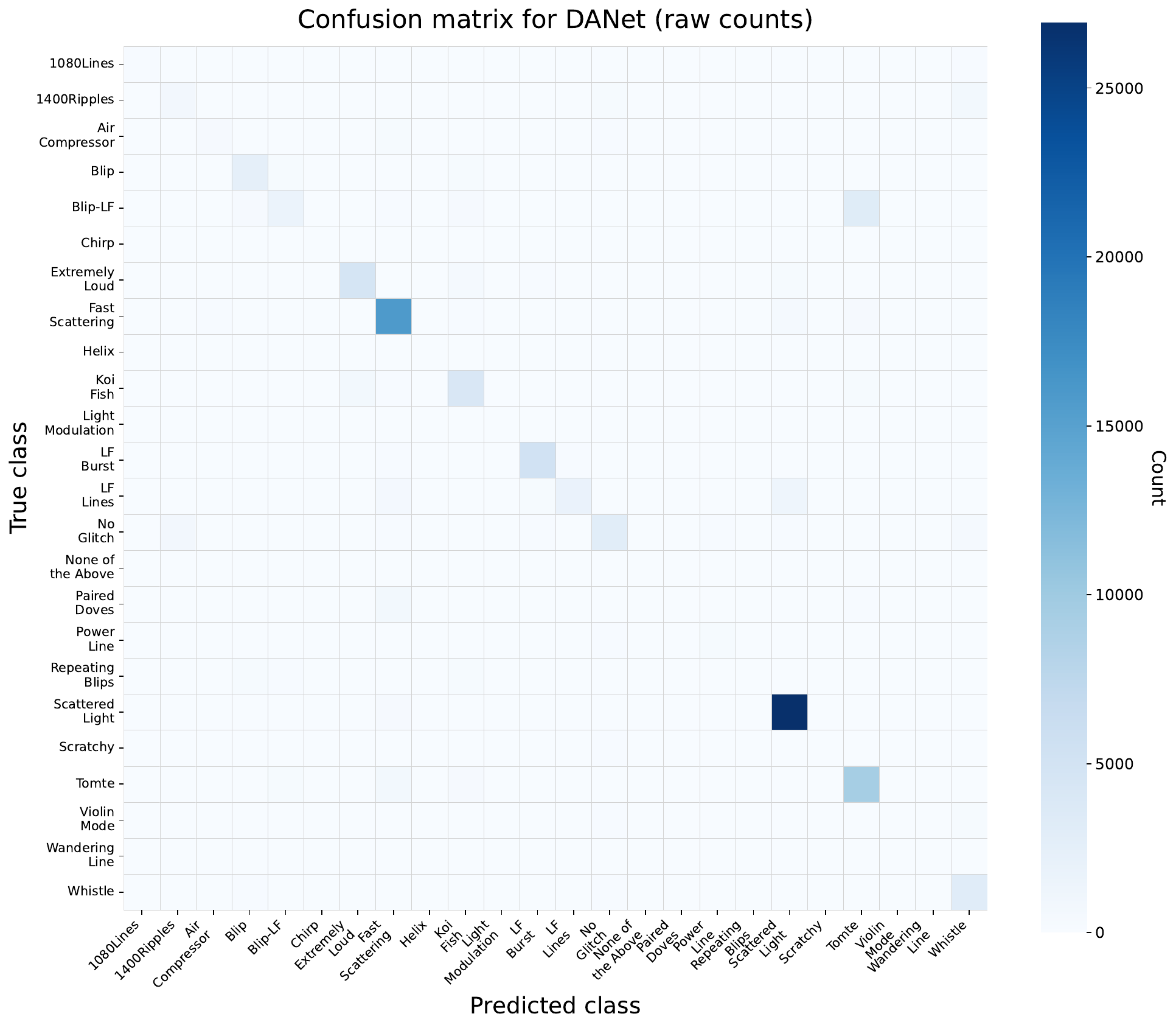}
\caption{\textbf{Confusion matrix for DANet showing raw prediction counts across the 24 glitch classes.} Each entry denotes the number of samples from a given true class assigned to a predicted class. The matrix reflects the underlying class distribution of the dataset, with dominant classes contributing larger counts along the diagonal. Consistent off-diagonal patterns corroborate systematic misclassifications observed in the normalized matrix and provide additional context on the absolute frequency of these errors.}
\label{fig:confusion_raw}
\end{figure}

The corresponding raw-count confusion matrix (Figure~\ref{fig:confusion_raw}) confirms that these patterns are not artifacts of normalization but persist across absolute sample counts. In addition, the raw matrix reveals substantial variation in class support, indicating that part of the observed performance differences across classes is driven by dataset imbalance.

This analysis provides insight beyond aggregate performance metrics, revealing that while DANet achieves strong overall accuracy, its ability to distinguish between certain physically similar glitch types remains limited. The persistence of structured confusion patterns across the full dataset suggests that these errors are systematic rather than due to statistical noise. These results highlight the inherent challenges of representing complex time–frequency structures using tabular features and motivate future work incorporating richer representations or domain-specific feature engineering.

\section{Discussion} \label{sec:discussion}
\noindent
The results presented in Figures~1–8 collectively demonstrate that performance, efficiency, model complexity, and interpretability represent largely orthogonal axes in tabular gravitational-wave glitch classification. While gradient-boosted decision trees remain a strong baseline in terms of absolute classification performance, several deep learning architectures achieve competitive accuracy with substantially fewer parameters and favorable inference-time characteristics. This highlights that raw model capacity alone is not a reliable predictor of empirical performance on structured detector metadata.

The robustness analysis in Figure~3 shows that high-performing models also exhibit stable behavior across random initializations, mitigating concerns that reported gains arise from favorable seeds or stochastic effects. Efficiency-oriented comparisons further clarify practical trade-offs: Figures~4 and~5 illustrate that training cost and inference latency differ by orders of magnitude across models, emphasizing that model choice should be guided by deployment constraints as much as by peak accuracy. In operational settings, such as online glitch classification during observing runs or real-time detector monitoring pipelines operating at fixed cadence, per-event inference latency at the millisecond scale may be desirable to ensure sustained throughput and minimal backlog. In such contexts, architectural differences that appear modest in offline benchmarking can become operationally significant when scaled to continuous data streams.

Importantly, the model-complexity analysis in Figure~6 provides a scientifically cleaner perspective on efficiency than training-time penalties, revealing regimes of parameter efficiency, saturation, and over-parameterization. In particular, several architectures achieve strong performance with orders-of-magnitude fewer trainable parameters than larger neural models, underscoring the role of inductive bias in tabular learning.

Beyond performance metrics, Figures~7 and~8 offer a detailed interpretability analysis that goes beyond prior work in gravitational-wave glitch classification. The one-to-one alignment with XGBoost demonstrates that certain deep learning models recover feature rankings consistent with trusted tree-based baselines, while the cross-model heatmap reveals clustering of feature-importance structure according to architectural principles. To our knowledge, this constitutes the first systematic cross-model interpretability alignment study applied to gravitational-wave detector metadata. The observed convergence of feature hierarchies across multiple deep tabular models strengthens confidence that learned representations capture meaningful detector and signal properties rather than model-specific artifacts.

Figures~9 and~10 extend this analysis by examining class-level prediction behavior through confusion matrices for the best-performing model, DANet. While aggregate metrics indicate strong overall performance, the normalized confusion matrix reveals that classification errors are not uniformly distributed across classes. Instead, misclassifications are concentrated among specific groups of glitch types, particularly those with similar time–frequency morphology or overlapping statistical features. For example, confusion between low-frequency transient classes and between variants of blip-like glitches suggests that the current tabular feature representation does not fully disentangle these physically related phenomena. At the same time, several dominant classes exhibit strong diagonal structure, indicating reliable separability when distinctive feature patterns are present. The corresponding raw-count matrix confirms that these trends persist in absolute terms and are not artifacts of normalization. Moreover, the raw matrix reveals substantial variation in class support, indicating that part of the observed performance differences across classes is driven by dataset imbalance.

Taken together, these findings clarify when deep learning models provide tangible advantages over classical approaches in tabular gravitational-wave analyses. Rather than advocating a wholesale replacement of established pipelines, our results support a more targeted adoption of deep learning, particularly in settings where parameter efficiency, interpretability alignment, or low-latency inference are prioritized. At the same time, the observed class-level confusions highlight an important limitation: improvements in model architecture alone may be insufficient to resolve ambiguities inherent to the feature space. The persistence of structured confusion patterns across the full dataset further indicates that these errors are systematic and physically meaningful, rather than arising from statistical noise. This motivates future work on richer representations and domain-informed feature engineering.

\section{Conclusions} \label{sec:conclusions}
\noindent
We have presented a systematic and controlled benchmark of classical and deep learning models for multiclass gravitational-wave glitch classification using tabular metadata from the Gravity Spy O3 dataset. By jointly evaluating classification performance, robustness, training and inference efficiency, model complexity, and interpretability alignment, we provide a comprehensive framework for model selection in detector characterization tasks.

Our results show that while tree-based ensembles remain highly effective, several deep learning architectures offer competitive accuracy with improved parameter efficiency and favorable deployment characteristics. Crucially, we demonstrate that interpretability structure can be quantitatively compared across models, revealing consistent feature hierarchies among distinct deep tabular architectures and highlighting fundamental differences between neural and tree-based approaches. This interpretability-focused analysis introduces a new dimension for evaluating machine-learning methods in gravitational-wave data analysis.

At the same time, class-level analysis reveals that certain morphologically similar glitch types remain difficult to distinguish using tabular features alone, indicating that improvements in model architecture may need to be complemented by richer representations of the underlying time–frequency structure. The persistence of structured confusion patterns across the full dataset further suggests that these errors are systematic and physically meaningful, rather than arising from statistical noise. In addition, variation in class support contributes to differences in per-class performance, highlighting the role of dataset imbalance in shaping model behavior.

Taken together, these results indicate that deep learning models provide targeted advantages rather than universal improvements over classical approaches, with benefits most pronounced in regimes prioritizing parameter efficiency, interpretability alignment, or deployment efficiency. 

Future work will extend this framework to hybrid representations that combine tabular metadata with time–frequency information, as well as to physics-informed architectures designed to incorporate detector knowledge directly into the learning process. More broadly, the methodology developed here provides a blueprint for principled, interpretable benchmarking of machine-learning models in gravitational-wave detector characterization and other data-intensive scientific applications.
\section*{Data Availability} \label{sec:data}
\noindent
The code, configuration files, summary results, and figure-generation scripts used to reproduce the results of this work are publicly available at the project GitHub repository: \url{https://github.com/rudhresh1997/glitch-tabular-prd}.

An archived, versioned release of the repository is available on Zenodo: \url{https://doi.org/10.5281/zenodo.19475319}.

The dataset used in this work is publicly available via Zenodo: \url{https://doi.org/10.5281/zenodo.5649212}.
\section*{Acknowledgements} \label{sec:acknowledgements}
\noindent
The author thanks the Gravity Spy collaboration for making the dataset publicly available. Computational resources for this work were provided by the Baylor University Kodiak High Performance Computing (HPC) facility.

The author is grateful to Dr.~Greg~Hamerly (Department of Computer Science, Baylor University) for valuable discussions and expert feedback on machine-learning methodology, and to Colin~Burdine (Department of Electrical and Computer Engineering, Baylor University) for helpful comments and insights that improved this work.

\nocite{*}

\appendix*
\section{Appendix A - Scaling with Data Volume} \label{sec:appendixa}
\noindent
To provide additional context on how different models respond to increased data availability, we examine performance changes when scaling the training set size from a sampled subset of approximately $50{,}000$ examples to the full Gravity Spy O3 dataset containing $\sim500{,}000$ examples. This analysis is presented as supplementary material, as it does not directly affect the primary performance–efficiency–interpretability conclusions discussed in the main text.

\begin{figure}[h]
\centering
\includegraphics[width=\linewidth]{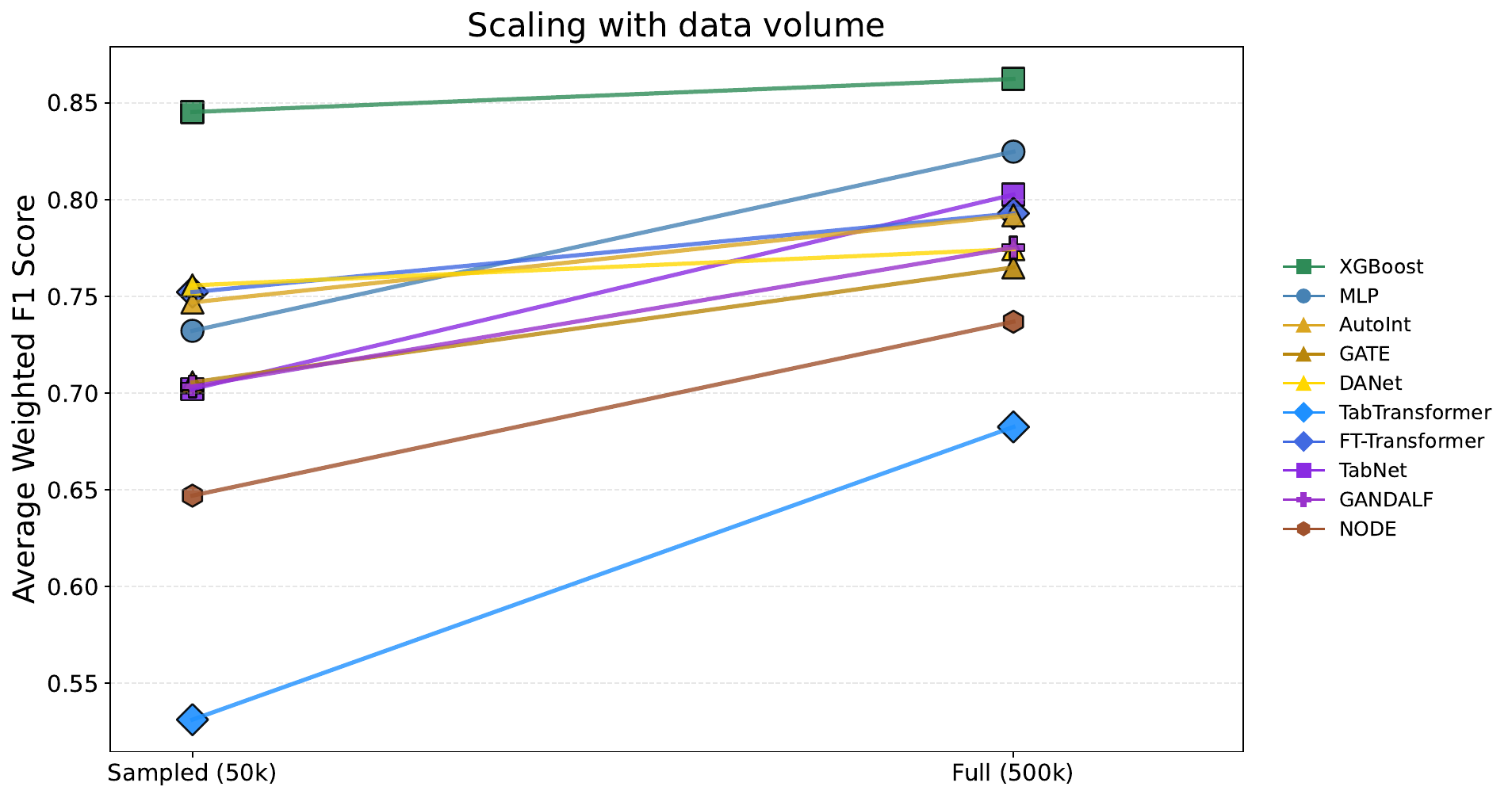}
\caption{\textbf{Scaling behavior of models as training data increases from $50{,}000$ to $\sim500{,}000$ examples.} Each line connects the weighted F1 score achieved on the sampled dataset to that obtained on the full dataset.}
\label{fig:scaling}
\end{figure}

Figure~\ref{fig:scaling} shows a slope plot of the weighted F1 score for each model evaluated on the sampled and full datasets. All models exhibit performance improvements with increased data volume, though the magnitude of the gain varies substantially across architectures. Neural models with lower initial performance on the sampled dataset—such as TabTransformer and NODE—benefit most strongly from additional data, while tree-based methods and well-regularized neural architectures show more modest but consistent gains.

Notably, the observed scaling behavior does not correlate monotonically with model complexity. Models with large parameter counts do not uniformly exhibit superior data efficiency, reinforcing the conclusion drawn from Figure~6 that inductive bias plays a more central role than raw capacity in tabular gravitational-wave classification. This appendix therefore supports, but does not extend, the primary conclusions of the paper by illustrating that increased data volume alone does not resolve architectural trade-offs.

\section{Appendix B - Global Feature Importance Across Models}
\label{sec:appendixb}
\noindent
To complement the cross-model interpretability analyses presented in the main text, we compute a global summary of feature importance by aggregating attribution scores across all models and glitch classes. For XGBoost, feature importances are derived from mean absolute TreeSHAP values, while for deep learning models we use mean absolute Integrated Gradients attributions computed with Captum. In all cases, feature importances are normalized within each model before aggregation to ensure comparability across explainers and architectures.

\begin{figure}[h]
\centering
\includegraphics[width=\linewidth]{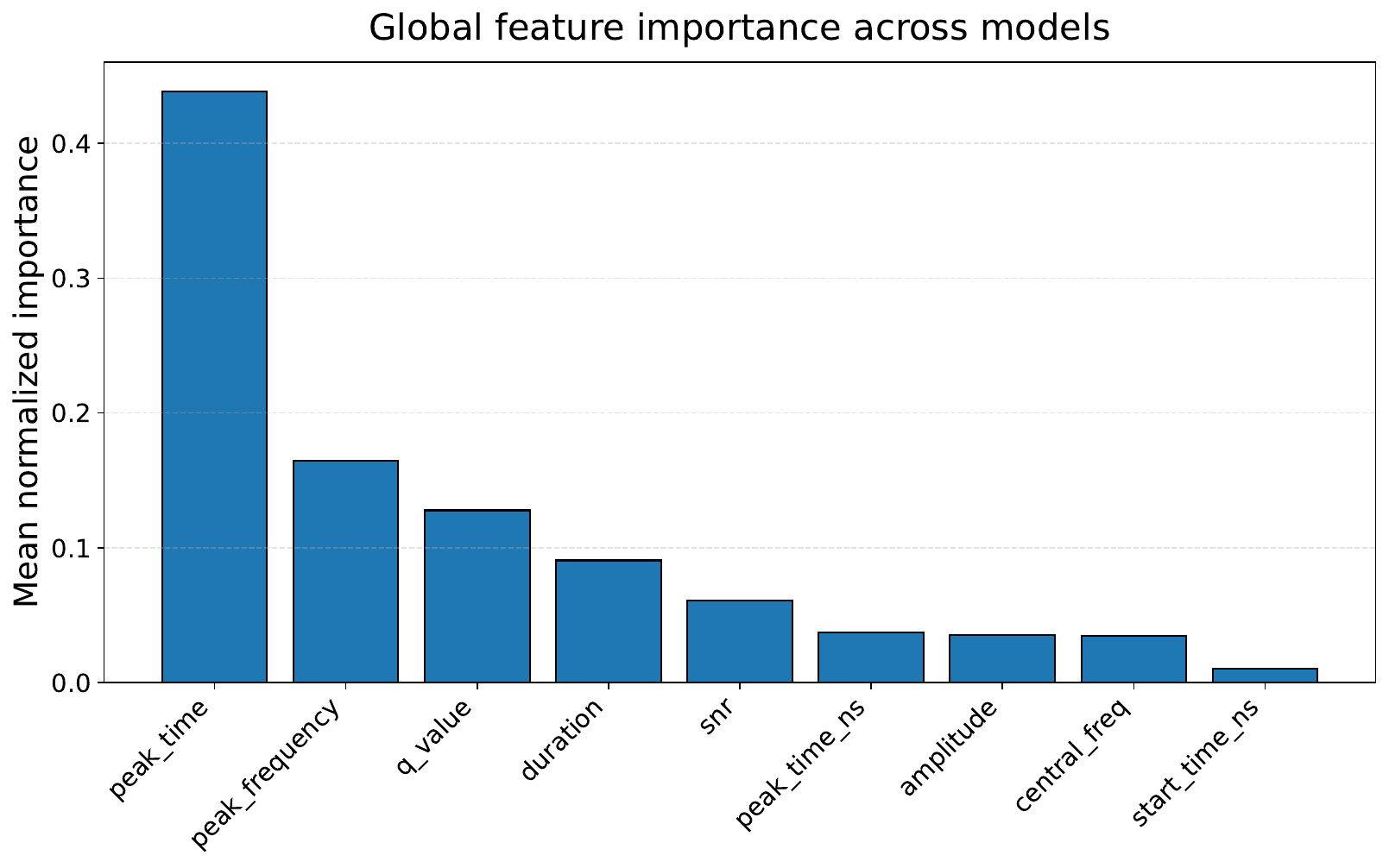}
\caption{\textbf{Global feature importance averaged across all models.} Feature importances are normalized within each model before aggregation.}
\label{fig:global_feature}
\end{figure}

Figure~\ref{fig:global_feature} shows the resulting global feature-importance ranking, averaged across models. A small subset of features dominates across architectures, with \texttt{peak\_time} and \texttt{peak\_frequency} emerging as the most influential variables, followed by \texttt{q\_value}, \texttt{duration}, and \texttt{snr}. Features related to fine-grained timing offsets and absolute amplitude play a comparatively smaller role when averaged across models.

This global view highlights robust, physically meaningful drivers of glitch classification that persist across diverse inductive biases and attribution methods. It provides an interpretable summary of which aspects of detector metadata are most consistently leveraged by machine-learning models in practice.

\section{Appendix C - Model-Wise Feature Importance Across Models}
\label{sec:appendixc}
\noindent
While Appendix B summarizes feature importance averaged across models, individual architectures may emphasize different aspects of the feature space. To expose these differences, we present model-wise feature-importance profiles for all evaluated models.

\begin{figure}[h]
\centering
\includegraphics[width=\linewidth]{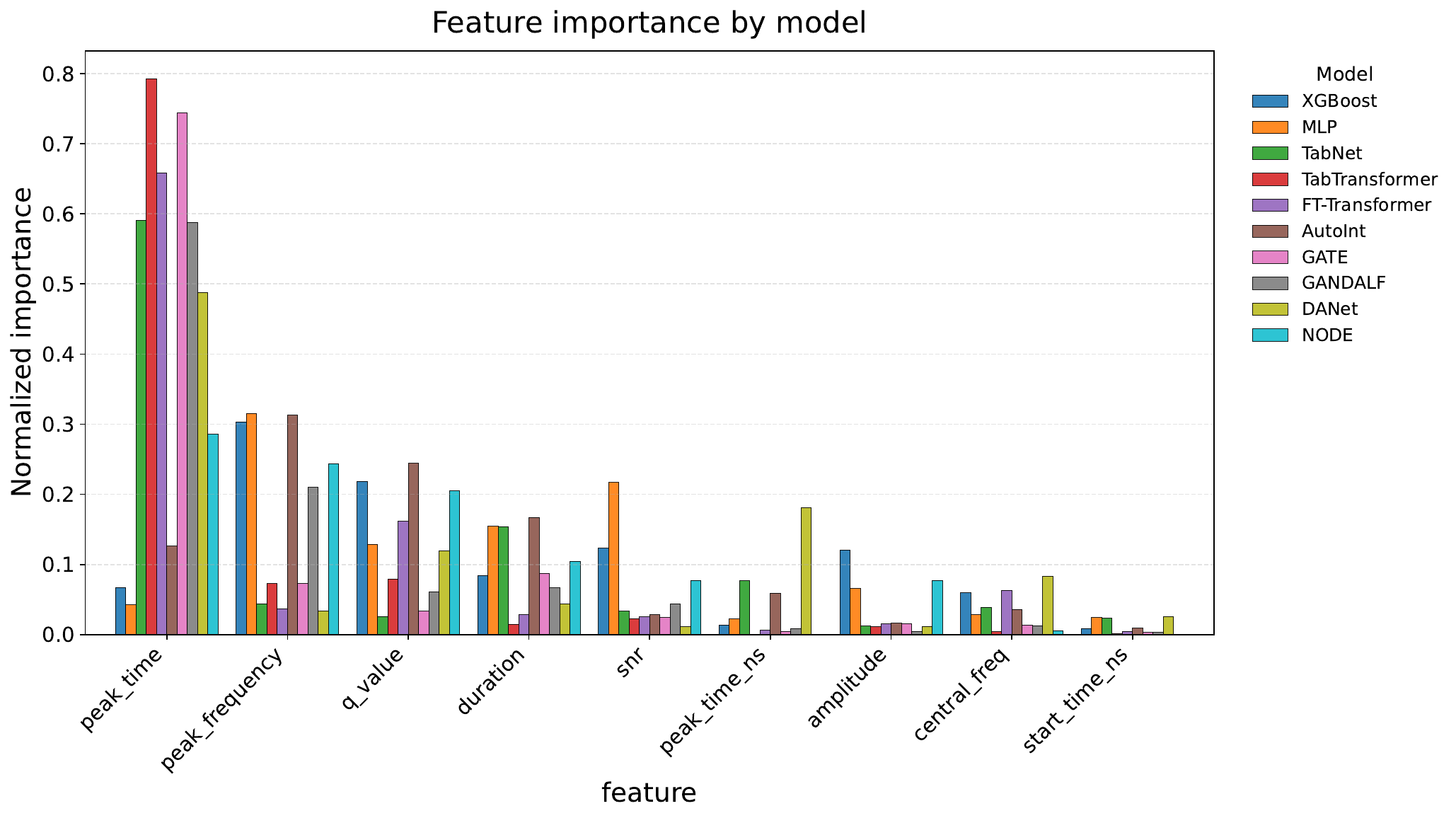}
\caption{\textbf{Normalized feature-importance profiles for each model.} Importances are computed using TreeSHAP for XGBoost and Integrated Gradients for deep learning models, then normalized within each model.}
\label{fig:modelwise_feature}
\end{figure}

Figure~\ref{fig:modelwise_feature} shows normalized feature importances for each model separately. Although broad qualitative agreement exists—particularly for dominant features such as \texttt{peak\_time} and \texttt{peak\_frequency}—substantial variation is evident in secondary features. For example, attention-based models tend to distribute importance more evenly across temporal and spectral descriptors, while tree-based and MLP models exhibit sharper concentration on a smaller subset of features.

These model-specific patterns help contextualize the cross-model alignment results discussed in Section~\ref{sec:results}. Together, the global and model-wise views demonstrate that interpretability agreement arises from shared emphasis on a common core of physically relevant features, even as architectural differences lead to distinct attribution signatures.

\end{document}